\documentclass{optica-article}

\journal{oe}

\articletype{Research Article}

\usepackage{lineno}
\usepackage{gensymb}
\usepackage{amsmath}
\begin{document}

\title{Polarization Independent Grating in GaN-on-Sapphire Photonic Integrated Circuit}

\author{Suraj\authormark{1}, Shashwat Rathkanthiwar\authormark{1,2}, Srinivasan Raghavan\authormark{1}, Shankar Kumar Selvaraja\authormark{1,*}}

\address{\authormark{1} Center for Nanoscience and Engineering, Indian Institute of Science, Bangalore-560012, India}

\address{\authormark{2} Now at the Department of Materials Science and Engineering,
North Carolina State University, USA.}

\email{\authormark{*}shankarks@iisc.ac.in} 



\begin{abstract}
In this work, we report the realization of a polarization-insensitive grating coupler, single-mode waveguide, and ring resonator in the GaN-on-Sapphire platform. We provide a detailed demonstration of the material characterization, device simulation, and experimental results. We achieve a grating coupler efficiency of -5.2 dB/coupler with a 1dB and 3dB bandwidth of 40 nm and 80 nm, respectively. We measure a single-mode waveguide loss of -6 dB/cm. The losses measured here are the lowest in a GaN-on-Sapphire photonic circuit. This demonstration provides opportunities for the development of on-chip linear and non-linear optical processes using the GaN-on-Sapphire platform. To the best of our knowledge, this is the first demonstration of an integrated photonic device using a GaN HEMT stack with 2D electron gas. 
\end{abstract}


\section{Introduction}

Gallium Nitride (GaN) has increasingly become promising in the field of photonics due to broadband transparency and the ability to grow high optical quality wafer-scale thin films. GaN with a typical bandgap of$\approx$3.4 eV \cite{amir2021review}, can be used for photonic applications in the UV region such as UV photodetectors, LEDs, distributed feedback (DFB) lasers \cite{dylewicz2004applications,bockowski2014bulk,cheng2021multicolor,ko2020hexagonal
,chang2006laser,smith2021high}. Further, the ease of bandgap engineering in GaN by alloying with other III-V nitrides such as InN and AlN\cite{dong2010band,cui2010band}, makes it applicable in UV, visible, to infrared photonics. Most of these applications utilize the bulk property of the GaN or its alloys. However, there are different ways to tailor GaN properties apart from alloying, which include doping \cite{eiting1998p}, quantum well-based bandgap engineering, and application of external electric field (Franz-Keldysh effect and Quantum confined stark effect (QCSE)) \cite{wetzel1999piezoelectric,maeda2018franz}. The inter-sub-band transition in quantum wells can be exploited to achieve a broader operating range from UV-IR. The GaN bandgap engineering enables control over the device functionality post-fabrication, leading to the realization of efficient active photonic devices. 

The interesting non-linear properties  such as Pockels and Kerr effect can help in fabricating GaN-based efficient optical modulators. The presence of a non-centrosymmetric structure of the GaN crystal enables the material to generate a second-order non-linear-optical response, such as a second harmonic generation from the bulk crystal. The presence of non-linear susceptibility $\chi$(2) of the order of lithium niobate (value approx. 20 pm/V) provides a platform that can be used to realize frequency conversion over the transparency window of GaN\cite{zheng2022integrated,xiong2011integrated,chen2017characterizations,
miragliotta1993linear}. Recently, there has been a demonstration of four-wave mixing in GaN/AlGaN stack \cite{munk2018four}. The advantage of implementing the non-linear optical activities at the thin film is that the power and the footprint required to achieve the desired non-linear activity reduces due to increased optical intensity\cite{Miragliotta:93}. Despite so many advantages of GaN, there is not adequate work on grating coupler for GaN waveguides that efficiently couples light from free space to achieve a photonic integrated circuit, especially at communication wavelengths (1550 nm) with one report on free-standing GaN grating\cite{liu2018free}.

Efficient on-chip integrated photonic devices require epitaxially grown GaN films that are substrate specific. However, epitaxial growth of GaN-on-Si (100) faces two major challenges, a large lattice constant mismatch (>17\%) and the difference in the coefficient of thermal expansion ($\approx$116\%) between GaN and Si. The lattice mismatch leads to defective film, and high thermal mismatch would result in film cracks \cite{ng2021group}. This lattice mismatch could be reduced by using buffer layers, however, the thermal mismatch is still an issue for thicker GaN-film growth. Substrates such as Sapphire and silicon carbide provide an excellent growth surface to grow GaN with a lattice mismatch of 16\% and 3.5\%, and a minimal thermal mismatch of 25\%. By tuning the growth parameters of the buffer layer, thicker GaN-film growth is possible for realizing photonic devices.\\
In this paper, we present GaN-on-Sapphire as a substrate for integrated photonics applications. In particular, we present high electron mobility (HEM) stack that can be used to realize high-frequency electro-optic devices. We demonstrate GaN waveguide, grating fiber-chip coupler, and ring resonator as building block devices for a GaN-based photonic circuit. Furthermore, we present a polarization-independent vertical fiber-chip coupler using one-dimensional grating for the first time. We validate the dual-polarization excitation using the spectral response of a microring resonator.

\section{GaN film growth and characterization}
\begin{figure}[htbp]
\centering\includegraphics[width=14cm]{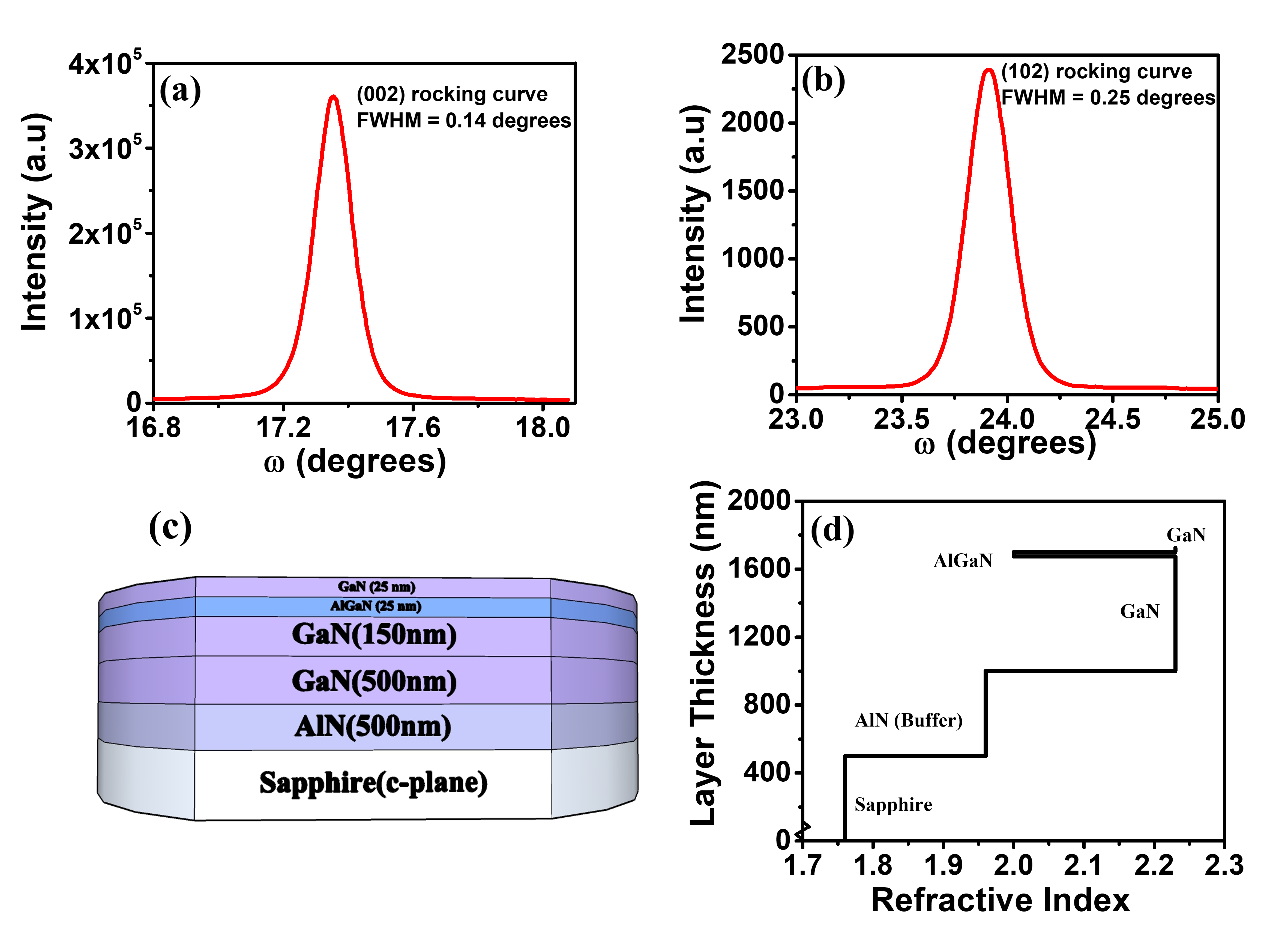}
\caption{(a),(b) The rocking $\omega$ scan of the (002) and (102) of the thin-film GaN. (c) The schematic of the high-electron mobility 2DEG stack used for the fabrication of waveguides and (d) the refractive index profile of the stack.}
\label{fig:material}
\end{figure}
Sapphire is chosen as the substrate to grow GaN of desired thickness. AlN is chosen as the intermediate layer between sapphire and GaN. AlN has a lattice mismatch of about 13.3\%\cite{TRAMPERT1997167,LIU200261} with Sapphire and hence gives a step transition from Sapphire to a GaN layer. Metal-organic chemical vapour deposition (MOCVD) was used to grow AlN and GaN on a sapphire substrate. Trimethylaluminum, Trimethylgalliumand, and NH$_3$ are used as Ga, Al, and N precursors with H$_2$ and N$_2$ as the carrier gases. The deposition and optimization of AlN and GaN are comprehensively studied in \cite{rathkanthiwar2020impact}. A 500 nm of AlN buffer layer is grown on Sapphire. GaN was deposited in two steps, initially 500 nm followed by 150 nm. The double-step growth leads to an epitaxial GaN film with sub-1 nm roughness with minimal defects \cite{rathkanthiwar2020impact}. The 2D electron gas (2DEG) layer is formed between the GaN and the AlGaN Layers, which is grown on top of GaN. The 2DEG can be exploited to realize electro-optic devices. Finally, 25 nm of GaN is grown on top of AlGaN as a passivation cap layer. Fig.\ref{fig:material} shows the characterization of the grown GaN on Sapphire with the seed layer and the resultant stack. The XRD $\omega$ scan data, as shown in Fig.\ref{fig:material}(a) \& (b), confirm epitaxial GaN growth. Small surface roughness leads to reduced scattering loss. The material stack used for photonic device fabrication is shown in Fig.\ref{fig:material}(c). Fig.\ref{fig:material}(d) shows the index profile along the stack. 
\begin{figure}[b]
\centering\includegraphics[width=12cm]{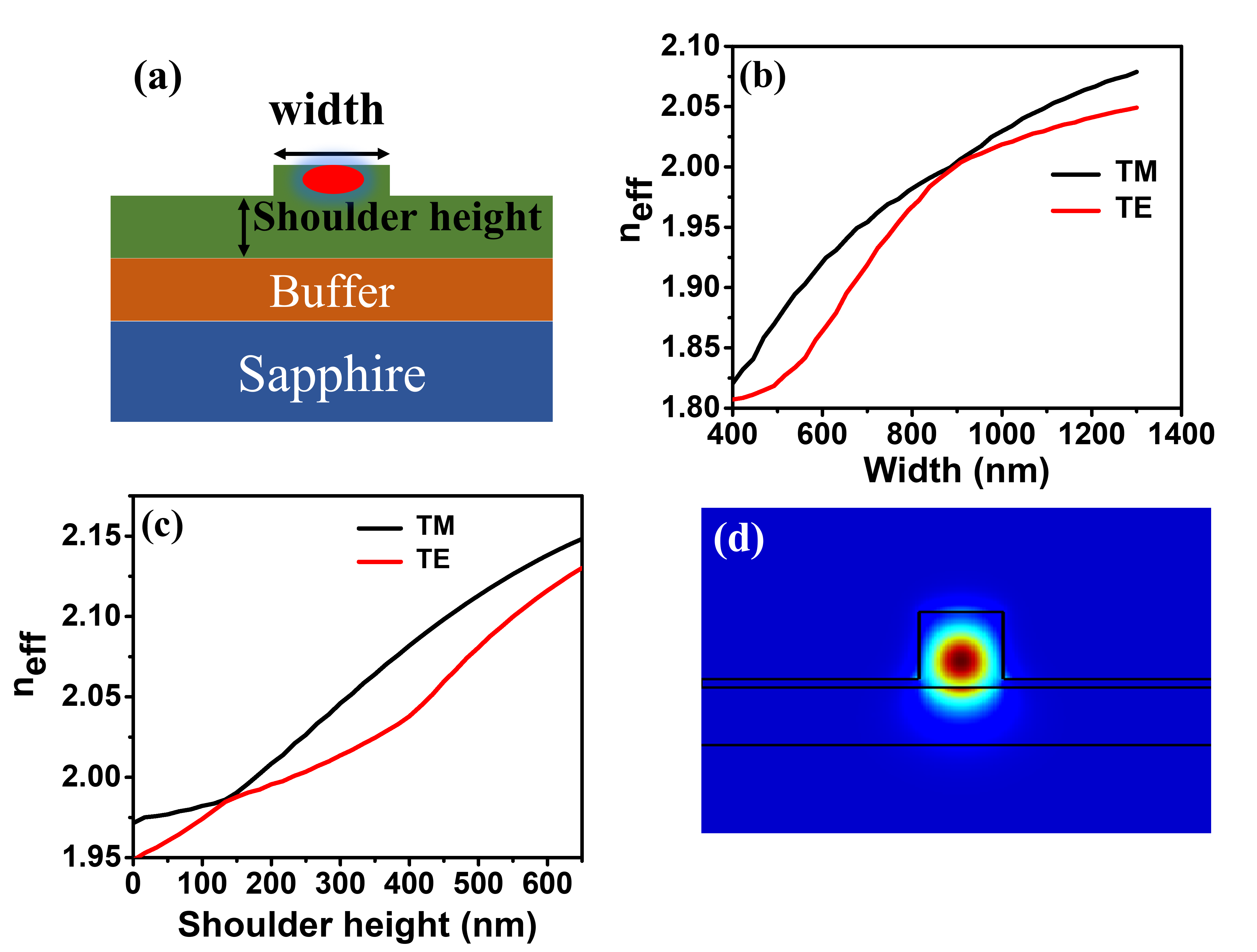}
\caption{(a)The schematic of the proposed GaN waveguide design.Waveguide dispersion with variation in the waveguide (b) Width (with shoulder height = 0 nm); (c) shoulder height (at width = 760 nm);(d) Mode field confinement at 760 nm waveguide width and 100 nm waveguide height.}
\label{fig:dispersion}
\end{figure}

\begin{figure}[htbp]
\centering\includegraphics[width=12cm]{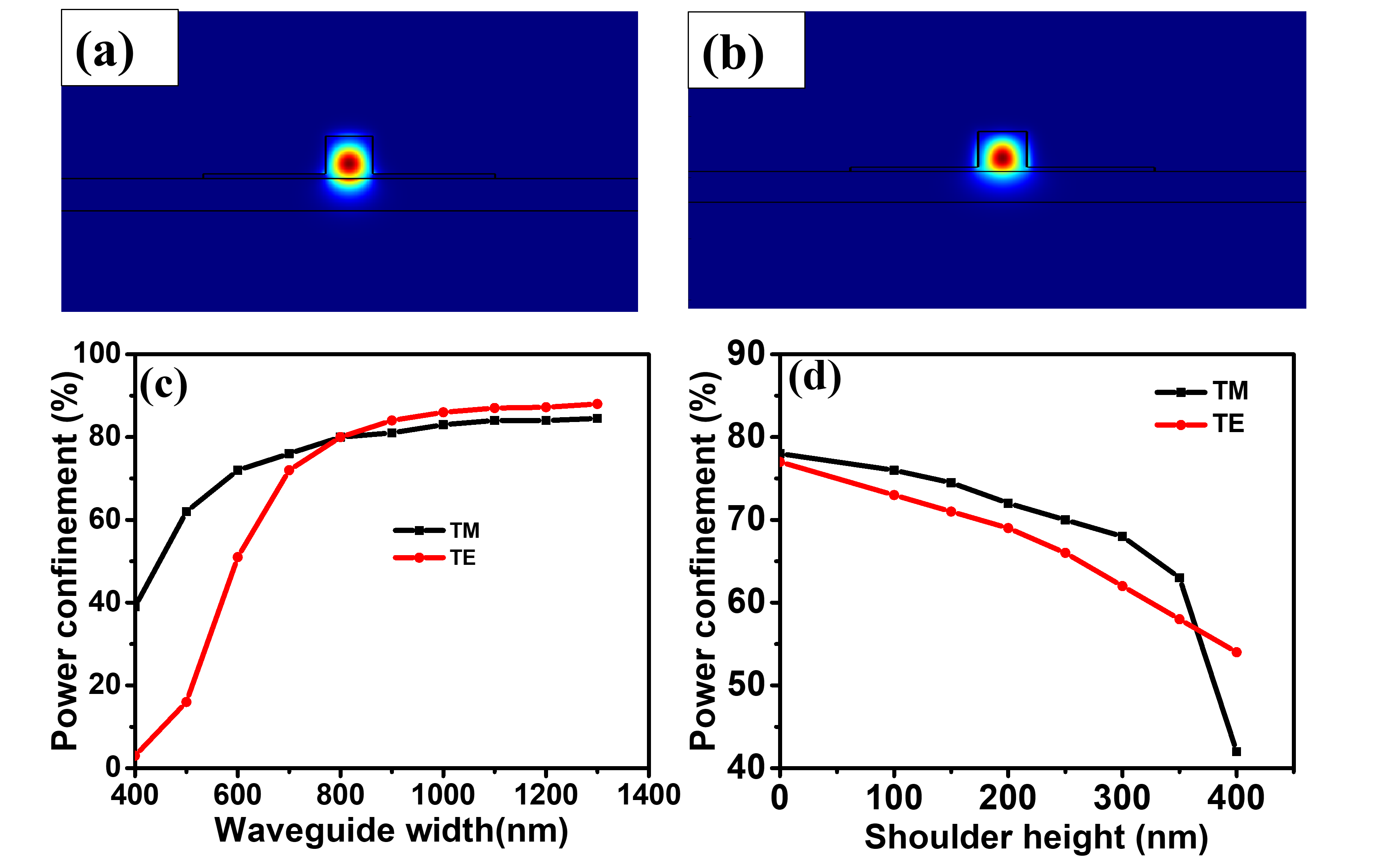}
\caption{Simulated power confinement with a waveguide dimension of 760 nm width and a shoulder height of 100 nm for (a)TM mode and; (b) TE mode; Variation of power confinement for TE and TM mode with varying (c) waveguide width (with shoulder height = 0 nm) and (d) shoulder height(at width = 760 nm).}
\label{fig:power_confinement}
\end{figure}

\section{Design and Simulation}
The material optimization is followed by waveguide design and simulation to determine the designed dimension for light confinement and propagation. In section \ref{ssec:num1}, we discuss the effect of waveguide width and etch depth on the modal characteristics at 1550 nm. Section \ref{ssec:num2} presents a detailed grating light-chip coupler design and analysis. The grating coupler simulation is performed to achieve maximum fiber-chip coupling to the GaN waveguide. 

\subsection{Waveguide design} \label{ssec:num1}
As evident from the index profile in Fig.\ref{fig:material}(d), an index contrast of 0.25 (with AlN buffer layer) would enable light confinement in the GaN layer. The schematic of the proposed GaN waveguide design on a Sapphire substrate showing a pictorial depiction of field confinement in the waveguide region is presented in Fig. \ref{fig:dispersion}(a). We use Ansys Lumerical MODE solution to optimize the waveguide parameters to obtain fundamental mode operation. The waveguide dispersion for fundamental TE and TM are obtained for variable width, and shoulder height is shown in Fig.\ref{fig:dispersion}(b \& c). For a width of 900 nm (shoulder height = 0 nm) and a shoulder height of 150 nm (width = 760 nm) of the waveguide, we observe a mode crossing between fundamental TE and TM mode. Fig.\ref{fig:dispersion}(b \& c) indicate that engineering the waveguide parameters could lead to a low birefringence between the fundamental TE and TM modes that can enable  polarization-independent operations in a GaN waveguide. Fig.\ref{fig:dispersion}(d) shows the field confinement of the TE fundamental mode for a 760 nm width waveguide with a 100 nm shoulder height.  Figure \ref{fig:power_confinement}(a) and (b) shows the mode field confinement of the fundamental TE and TM mode in the waveguide. As is evident from Fig.\ref{fig:power_confinement}(c) and (d) that more than 70 \% of the power is confined in the waveguide for a waveguide width of more than 760 nm and shoulder height of less than 100 nm for both TE and TM modes. In Fig.\ref{fig:power_confinement}(d), the waveguide starts to become multi-moded, and power is distributed across the shoulder leading to a reduction in power confinement in the specified area.

\subsection{Grating coupler design simulation} \label{ssec:num2}

\begin{figure}[htbp]
\centering\includegraphics[width=13cm]{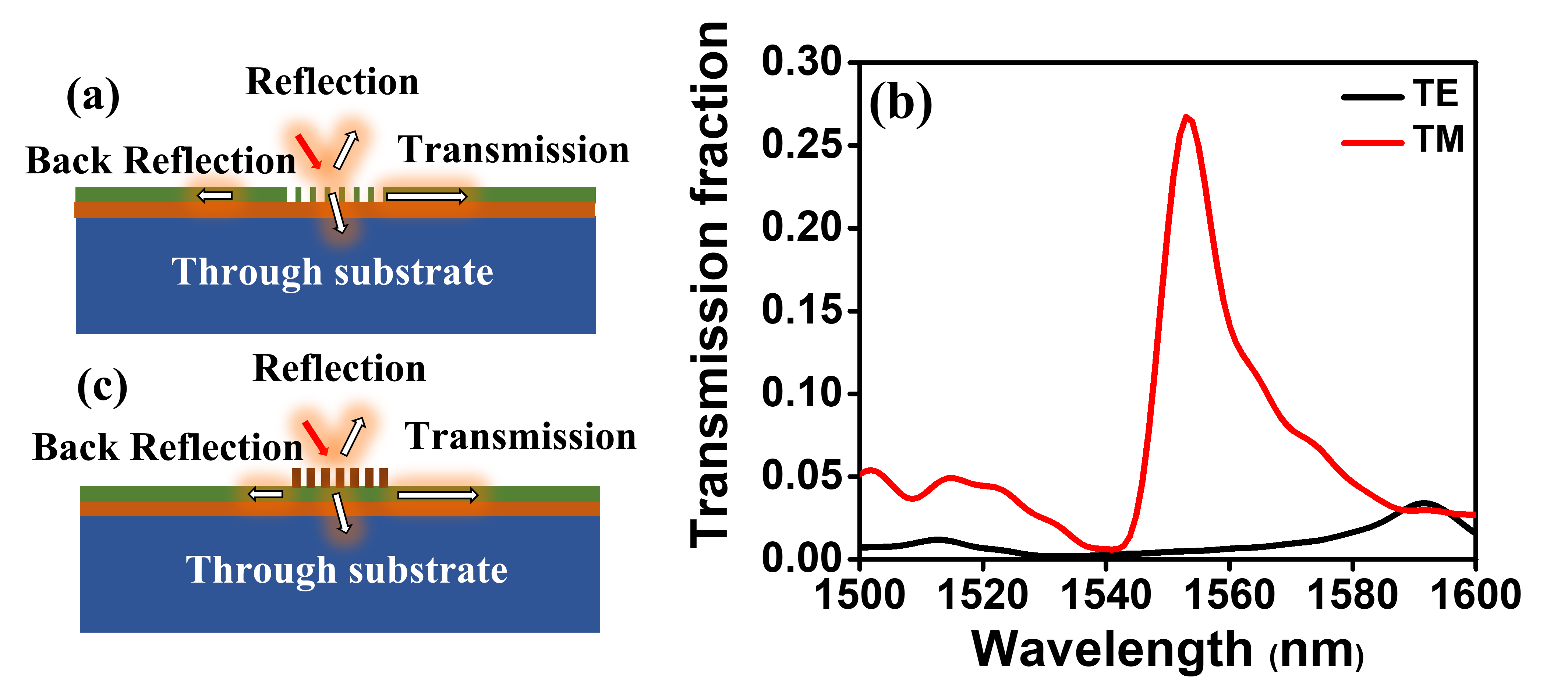}
\caption{(a) Schematic for GaN grating;(b) Coupling efficiency(CE) of GaN grating; (c) Schematic for a-Si overlay grating.}
\label{fig:GaN_grating}
\end{figure}

On-chip coupling can be performed in two ways; edge coupling and vertical/grating coupling. The devices need to be diced with optical quality chip-edge for high-efficiency coupling in edge coupling. Since sapphire is a hard material, polishing is a challenge. Furthermore, the alignment tolerance and ease of performing measurement is a challenge. In contrast, a grating coupler is relatively simple to couple light into a waveguide without any additional post-processing. A detailed design procedure is presented in \cite{app8071142}. Simulations were performed to determine the feasibility of the grating coupler on the proposed GaN stack.  Fig.\ref{fig:GaN_grating}(a) shows the schematic of the structure used to perform the mode simulations for the GaN gratings and Table \ref{tab:simulation} lists the refractive indices used for the various materials in the stack. Fig.\ref{fig:GaN_grating}(b) shows the coupling efficiency(CE) for a TE and TM mode using a Gaussian illumination.  For the GaN gratings, the TM mode CE is $\approx$ 30\% while the TE mode CE is $\approx$ 5\% indicating the polarization-sensitive nature of the grating coupler shown in (Fig.\ref{fig:GaN_grating}(b)).


\begin{table}[]
\centering
\caption{Refractive Index used in the simulation}
\label{tab:simulation}
\begin{tabular}
{|c|c|c|}
\hline
Sl. No & Material & Refractive Index (at 1550 nm) \\
\hline
1     & Sapphire & 1.76             \\
\hline
2     & AlN      & 1.96             \\
\hline
3     & GaN      & 2.23     
\\
\hline
4     & Silicon  & 3.43     
\\
\hline
\end{tabular}
\end{table}

One of the major challenges in coupling light into GaN is the low-index contrast as well as the absence of a bottom reflecting surface, such as the bottom substrate in a silicon-on-insulator. Since the light extraction is directly proportional to the light diffraction efficiency, the grating strength can be improved by adding a high-index overlay such as silicon \cite{roelkens2006high}. we choose amorphous silicon (a-Si) as the overlay grating material to simulate gratings on the GaN waveguide. The modified proposed design using Si grating on GaN waveguide to improve CE is shown in Fig.\ref{fig:GaN_grating}(c).

\begin{figure}[t]
\centering\includegraphics[width=14.5cm]{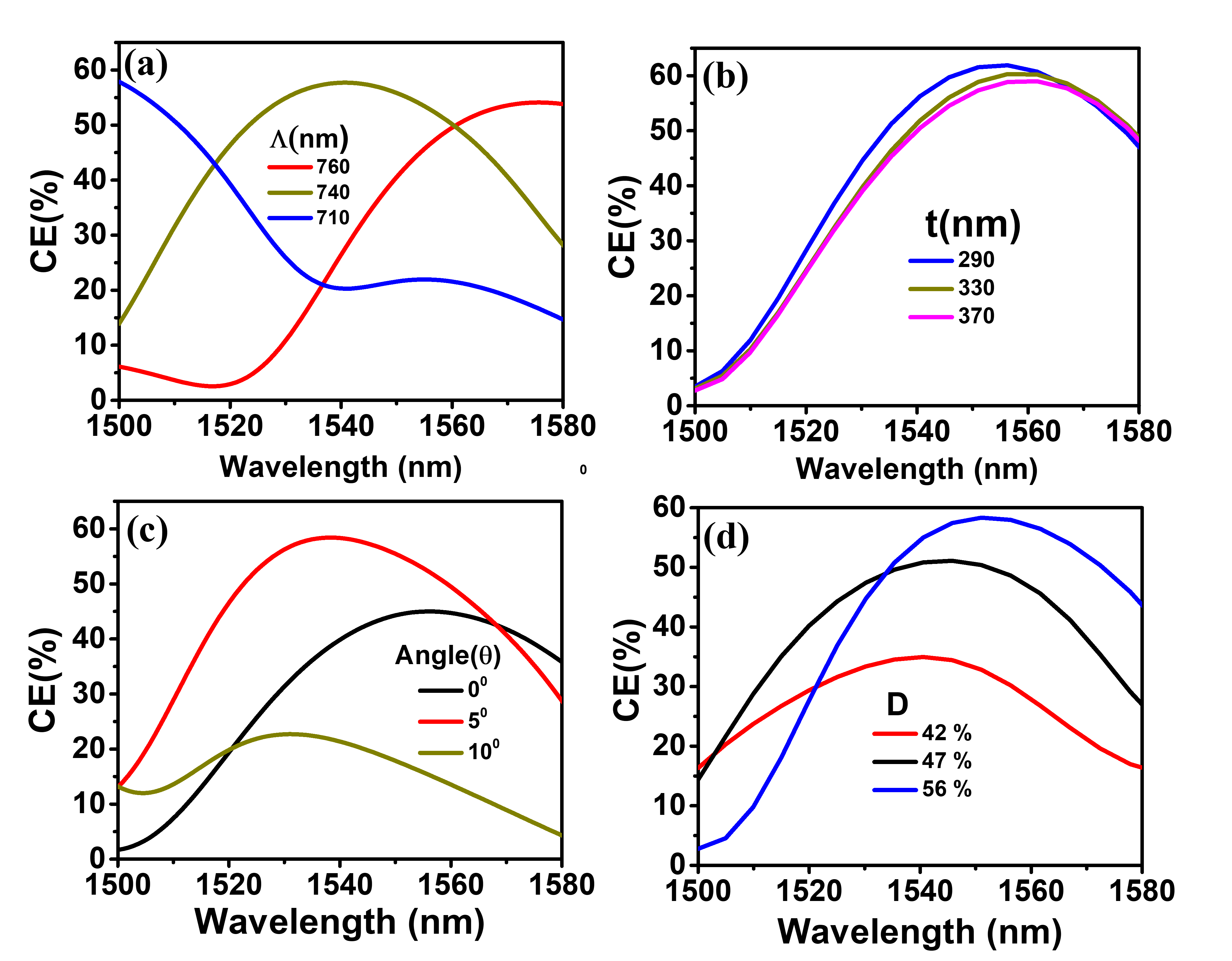}
\caption{Variation of Si grating coupler CE with (a) grating period(t=267 nm,$\theta$= 2\degree, grating width = 400 nm), (b) grating height($\Lambda$=744 nm,$\theta$= 2\degree,D = 54\%), (c) source angle($\Lambda$=744 nm,t= 267 nm,D = 54\%); and (d) Duty cycle($\Lambda$=744 nm,$\theta$= 2\degree,t = 267 nm) for TE mode.}
\label{fig:TE_grating_optimization}
\end{figure}

\begin{figure}[t]
\centering\includegraphics[width=14.5cm]{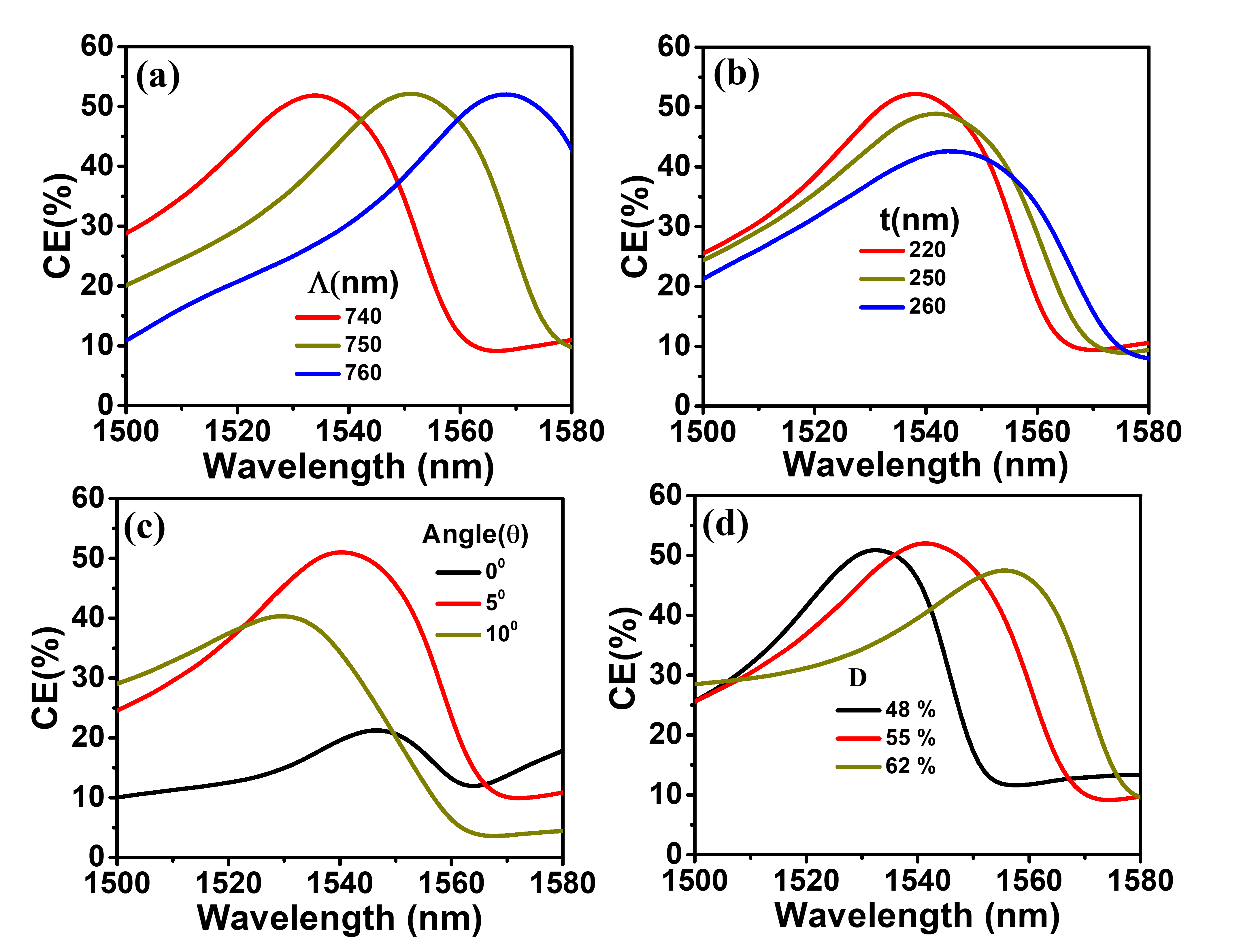}
\caption{Variation of Si grating coupler CE with (a) grating period(t=230 nm,$\theta$= 5\degree, grating width = 400 nm), (b) grating height($\Lambda$=744 nm,$\theta$= 5\degree, D = 54\%), (c) source angle($\Lambda$=744 nm,t= 230 nm,D = 54\%); and (d) Duty cycle($\Lambda$=744 nm,$\theta$= 5\degree,t = 230 nm) for TM mode.}
\label{fig:TM_grating_optimization}
\end{figure}

Implementation of Si-overlay grating increases the index contrast to 1.25 (360\% improvement from GaN-only grating) at 1550 nm wavelength, which would lead to increased efficiency of light coupling into the GaN waveguide. 
Si grating coupler parameters such as periodicity ($\Lambda$), etch depth(t), duty cycle(D), and the angle of incidence($\theta$) were optimized using Ansys Lumerical FDTD for both TE and TM polarization. Fig.\ref{fig:TE_grating_optimization}(a-d) shows the variation of the CE of TE mode with respect to the grating parameters, and Fig.\ref{fig:TM_grating_optimization}(a-d) shows the variation for a TM mode. 
The maximum efficiency for coupling achieved for a TE gaussian source with Si grating is approximately 60 \%, while for a TM gaussian source, it is 50 \%. The grating parameters for the maximum CE for a TE source are a grating period of 725 nm, an overlay height of 300 nm, and a duty cycle of 52.4\% with a source angle of 2\degree. The grating parameter for maximum CE for TM gaussian is a period of 750 nm, an overlay height of 260 nm, and a duty cycle of 48\% with an angle of 5\degree. As is observed in  Fig.\ref{fig:TE_grating_optimization}(a \& d) and Fig.\ref{fig:TM_grating_optimization}(a \& d), there is a red shift in the spectrum with increasing $\Lambda$ and duty cycle (D) for both TE and TM mode. However, we find that the coupler wavelength and efficiency are tolerant to the overlay thickness evident in Fig.\ref{fig:TE_grating_optimization}(b) and Fig.\ref{fig:TM_grating_optimization}(b). Though the spectral shift due to incident angle is small, the coupler efficiency is sensitive to incident angle variation as seen in Fig.\ref{fig:TE_grating_optimization}(c) and Fig.\ref{fig:TM_grating_optimization}(c).  
 The variation of TE grating CE with respect to the periodicity, etch depth, angle of incidence, and duty cycle are 1.04\% / nm, 0.36\% / nm, 5.62\% / degree, and 1.2\% / (change in D). The corresponding sensitivity for TM mode is  1.01\% / nm, 0.30\% / nm, 4.51\% / degree, 1.16\% / (change in D).

\begin{figure}[htbp]
\centering\includegraphics[width=7.5cm]{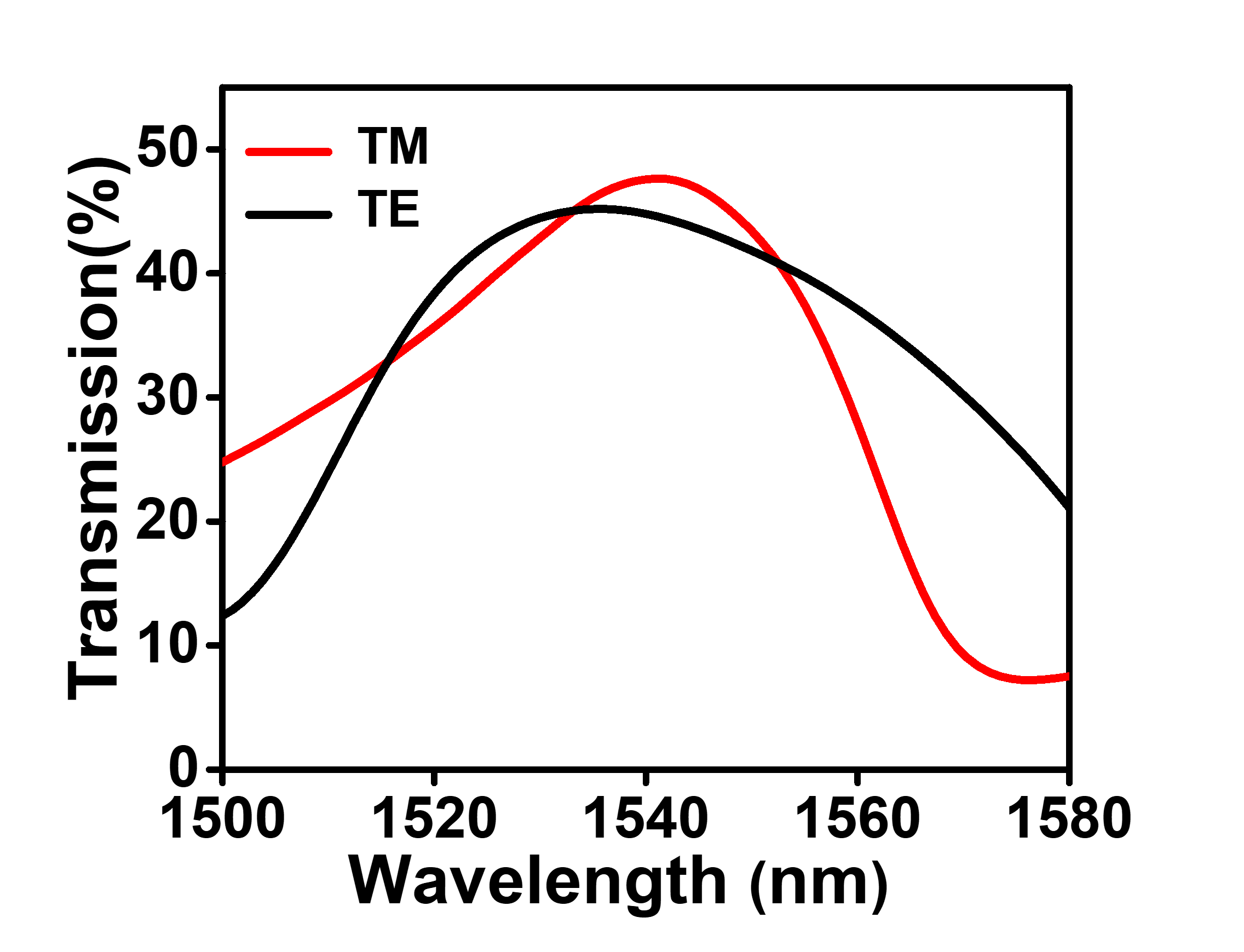}
\caption{Simulated transmission characteristic of Si grating coupler demonstrating polarization independent characteristic ($\Lambda$=750 nm,$\theta$= 5\degree, D = 48\%,t=260 nm).}
\label{fig:Polarization_compare}
\end{figure}

The results in  Fig.\ref{fig:TM_grating_optimization} indicate a maximum CE of $\approx$48\% for TM mode at a period of 750 nm, a grating height of 260 nm with a duty cycle of 48\% at 5\degree incident angle. For the same set of parameters, we obtained a CE of $\approx$48\% for a TE-polarized light. This suggests GaN waveguide with an engineered Si grating coupler could enable polarization-independent coupling characteristics. This is evident from Fig.\ref{fig:Polarization_compare}.
\begin{figure}[htbp]
\centering\includegraphics[width=14cm]{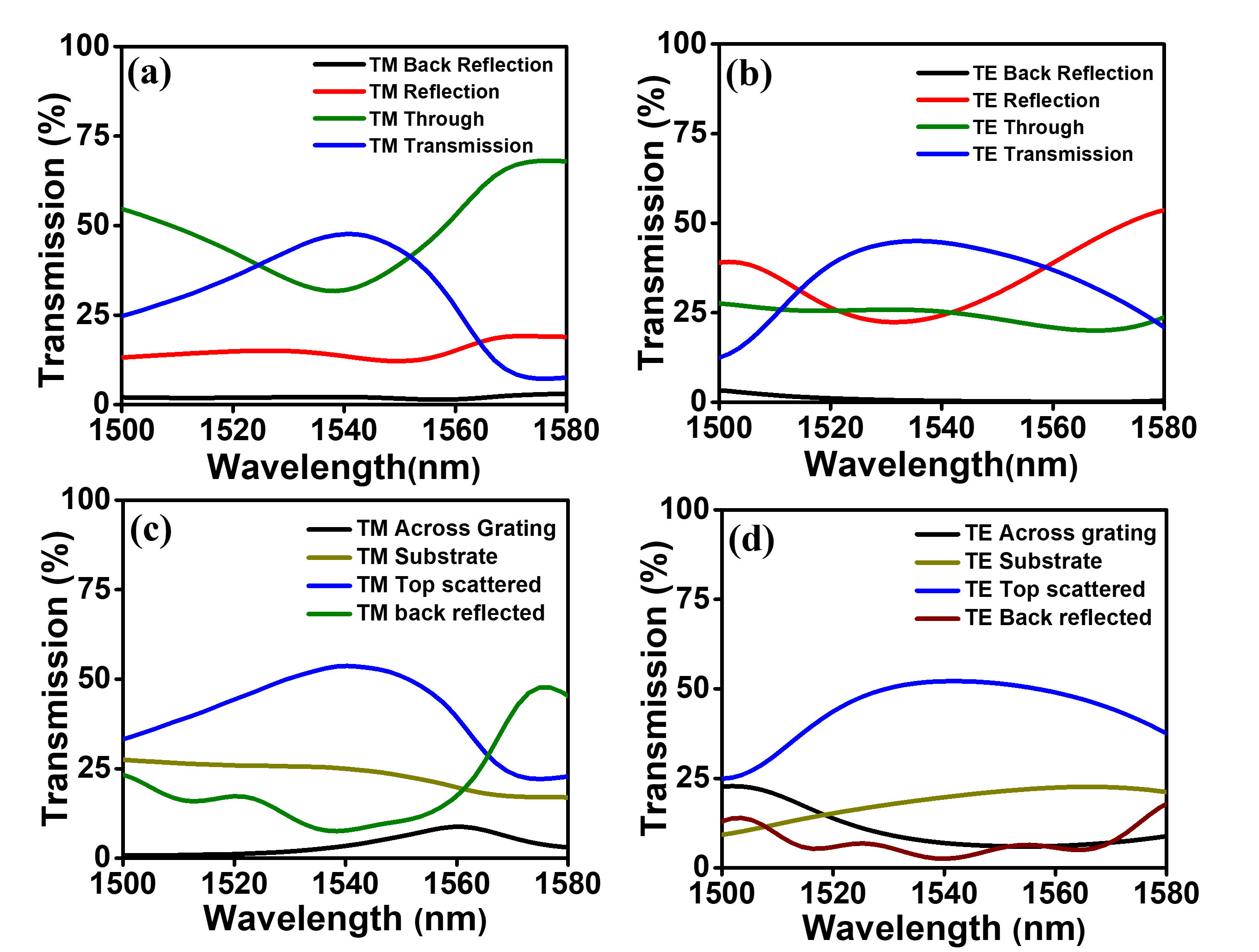}
\caption{ In-coupling simulation ($\Lambda$=750 nm,$\theta$= 5\degree, D = 48\%,t=260 nm) for (a) TM gaussian source  and (b) TE gaussian source; out-coupling simulation ($\Lambda$=750 nm,$\theta$= 5\degree, D = 48\%,t=260 nm) for (c) TM gaussian source and (d) TE Gaussian source.}
\label{fig:Loss_mechanism}
\end{figure}

Further, we simulate the coupling loss mechanism constituting the reflection, transmission, and substrate leakage of the proposed waveguide-grating coupler design for the optimized grating parameters. Fig.\ref{fig:Loss_mechanism}(a,b) show the in-coupling and Fig.\ref{fig:Loss_mechanism}(c,d) show the out-coupling power distribution. As seen in Fig.\ref{fig:Loss_mechanism}(a-d), substrate leakage is a major source of power loss for both TE and TM mode.

\section{Test Device fabrication}

\begin{figure}[htbp]
\centering\includegraphics[width=12cm]{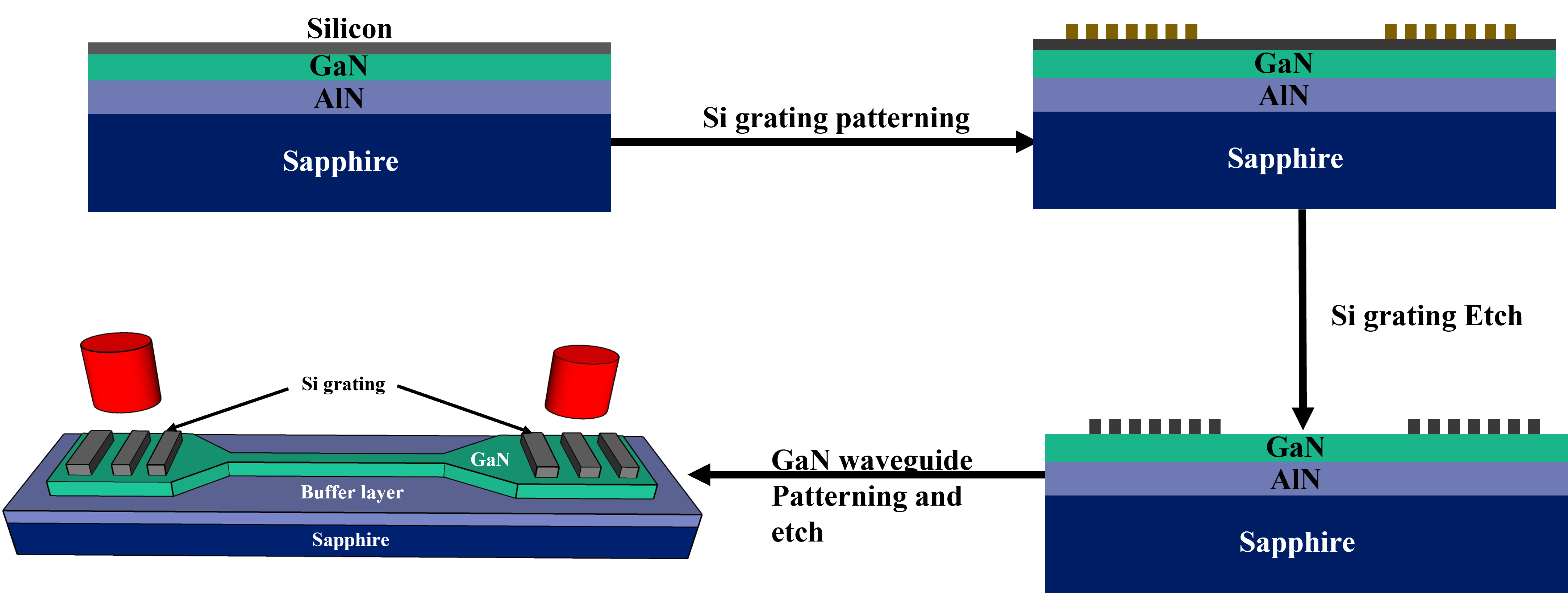}
\caption{Schematic for device fabrication process flow.}
\label{fig:Process_flow}
\end{figure}

Fig.\ref{fig:Process_flow} shows the schematic of the process flow used to fabricate GaN device. The GaN stack grown using MOCVD is coated with 235 nm of a-Si deposition. The deposition is followed by 15 nm of doped a-Si to reduce charging while performing electron-beam (e-beam) lithography. The pattering of gratings and GaN waveguide is done using e-beam lithography and dry etch process. 
500 $\mu m$ long patch waveguides of 10 $\mu m$ width and grating in and out couplers with various periods are fabricated as coupler test devices. We have also fabricated a microring resonator with a bend radius of 84 $\mu m$ and a coupling gap of 400 nm. 

\begin{figure}[htbp]
\centering\includegraphics[width=14cm]{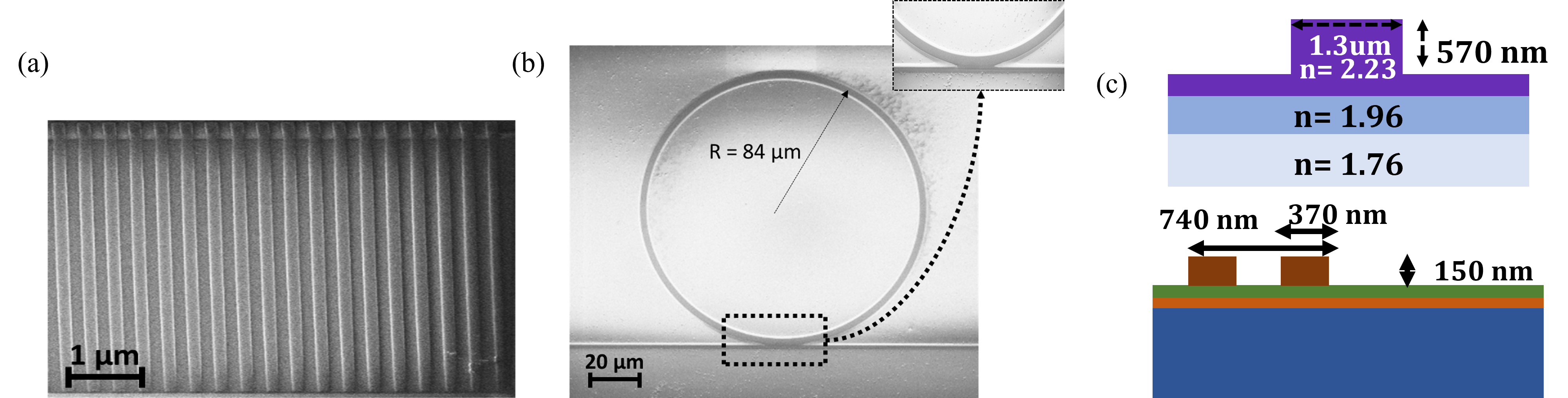}
\caption{FESEM of (a) grating and (b) ring,(c) stack based on refractive index and measurement based on AFM.}
\label{fig:Imaging}
\end{figure}

Fig.\ref{fig:Imaging}(a) and (b) show the field emission scanning electron microscopy(FESEM) images of the fabricated a-Si grating and GaN ring. Fig.\ref{fig:Imaging}(c) shows the schematic obtained post-AFM characterization. The fabricated dimensions were measured to be different from the desired design dimension due to process variation. The targeted parameters were a grating period of 750 nm, a grating height of 260 nm, and a duty cycle of 48\%. The fabricated device had a grating period of 740 nm, a grating height of 150 nm, and a waveguide etch depth of 570 nm instead of 650 nm. 
We observed that the roughness of the top GaN layer was around 20 nm. This is primarily due to dry etch selectivity between a-Si and GaN. The roughness could be improved by using a liner layer such as silicon dioxide or improved selectivity between GaN and a-Si. 

\begin{figure}[b]
\centering\includegraphics[width=14cm]{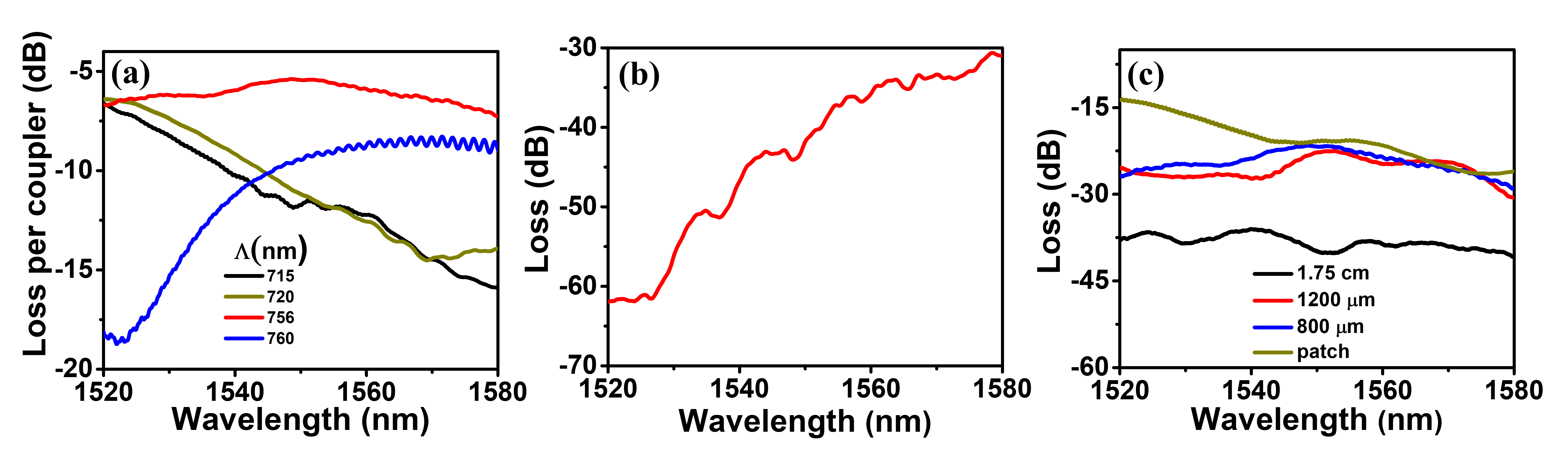}
\caption{Characterization for waveguide (a) Patch ; (b) spiral ;(c)loss calculation.}
\label{fig:Patch}
\end{figure}
\begin{figure}[t]
\centering\includegraphics[width=13.5cm]{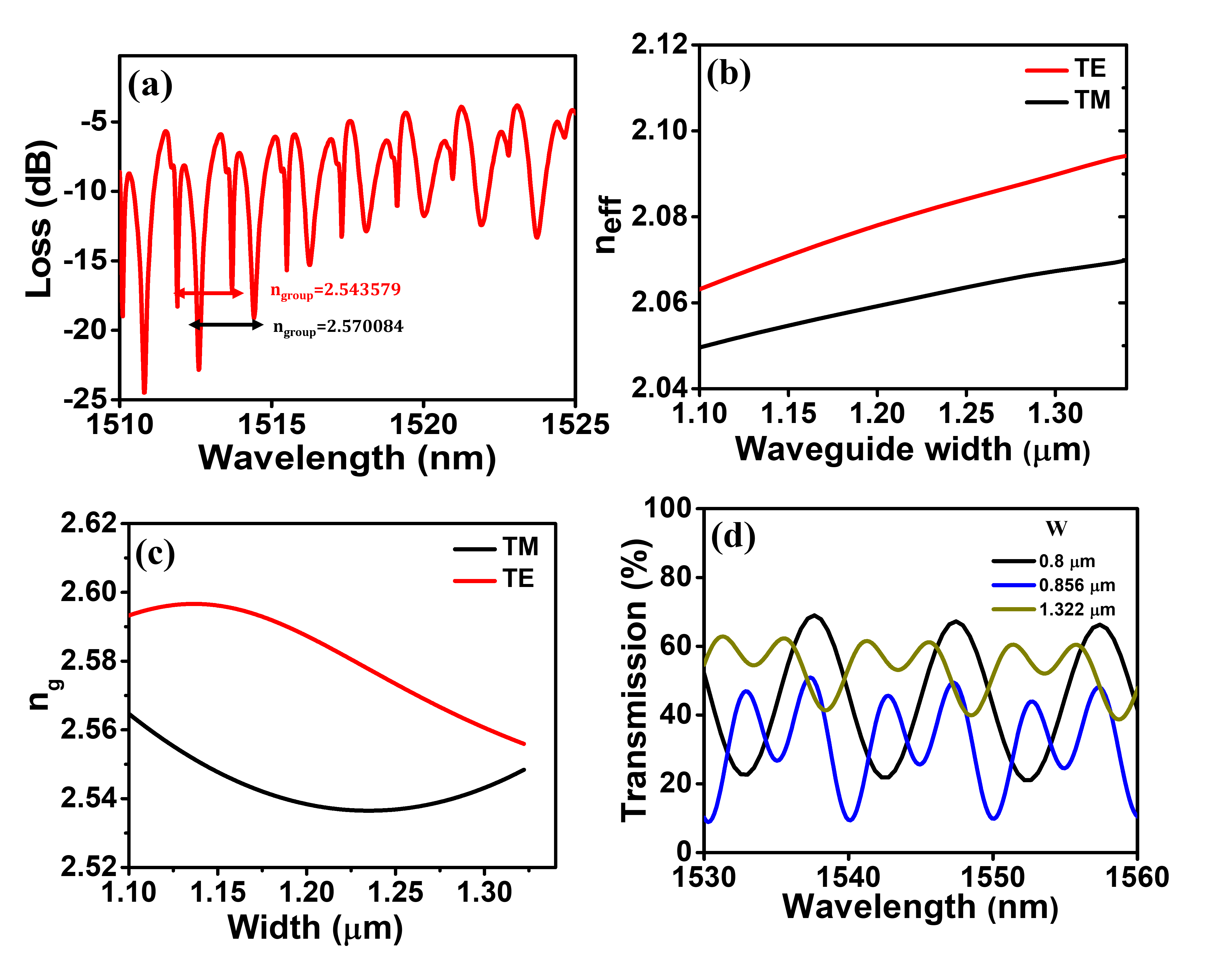}
\caption{(a) Measured Ring response; Simulated (b)waveguide dispersion; (c) waveguide group index variation; (d)ring characteristic for varying waveguide width.}
\label{fig:ring}
\end{figure}

\section{Characterization results summary}
Fig.\ref{fig:Patch} shows the characterization of the Si grating coupler, tapered and spiral waveguides using a tunable laser source. Fig.\ref{fig:Patch}(a) shows the result for a patch waveguide. We measure a CE of -5.2 dB from the fabricated device for a period of 756 nm. Fig.\ref{fig:Patch}(b) depicts the optical loss for a 1.75 cm long spiral waveguide. A fibre-to-fibre loss of -30.4 dB is measured. Fig.\ref{fig:Patch}(c) quantifies a propagation loss of 6.26 dB/cm for a single-mode 800 nm wide and 570 nm etched waveguide. We attribute the waveguide loss to the surface roughness resulting from a-Si etch and the corresponding side-wall roughness. The surface roughness induced by the a-Si is the primary cause of high propagation loss.
The output characteristics of the fabricated ring resonator are presented in Fig.\ref{fig:ring}(a). Due to small birefringence and similar TE and TM grating behavior, ring response was used to confirm the polarization-independent behavior. The Q-factor for the TE resonant mode obtained is 11,000, while for TM is 3,000.

The polarization-insensitive behavior of the gratings is further corroborated by simulating the waveguide dispersion, and group index variation of the propagating modes. The measured group index obtained from the two resonant modes is n$_g$=2.57 and 2.54 as shown in Fig.\ref{fig:ring}(a). Fig.\ref{fig:ring}(b) shows the waveguide dispersion with varying waveguide width with an etch depth of 570 nm. The propagation mode cut-off obtained for the fabricated device is 1.82. We observe for a width of 1.3 um, the waveguide supports a TE mode with a group index of 2.55 and TM mode with a group index of 2.54 that are comparable with the measured ones(2.57 and 2.54 respectively) as seen in Fig.\ref{fig:ring}(c). Further, we confirm the multimodal nature of the fabricated rings by simulating their output response at different widths. Fig.\ref{fig:ring}(d) shows that beyond 0.8 um, the ring-resonators showed multimodal behavior (multiple resonance dips) evident from their output characteristic. The multimodal feature, the ring responses, along with the measured group indices confirms the polarization-independent characteristic of the grating coupler fabricated on the GaN-sapphire platform.

\section{Conclusion}
In summary, for the first time, we demonstrated an efficient grating coupler in GaN waveguide. We presented a detailed simulation and fabrication process. We achieved a CE of -5.2 dB/coupler and a waveguide loss of $\approx$6 dB/cm. A 1dB bandwidth of 40 nm and a 3dB bandwidth of over 80 nm were obtained for the grating coupler. In addition, we employed a microring resonator, with a q-factor of 11,000 and 3,000 for TE and TM polarization, to demonstrate dual-polarisation coupling by the grating coupler demonstrating polarization-independent coupling by a single one-dimensional grating. To the best of our knowledge, this is the first such demonstration. The results presented are promising, with the scope to improve the performance through an optimized device fabrication process. This would open opportunities for using GaN as a waveguide material for linear and non-linear processing. 

\section{Acknowledgement}

We acknowledge funding from the Ministry of Education, Government of India, for supporting facilities at the Centre for Nanoscience and Engineering (CeNSE), Indian Institute of Science, Bangalore. SKS thanks Professor Ramakrishna Rao chair fellowship.

\section{Disclosures}
The authors declare no conflicts of interest.

\section{Data availability}
Data underlying the results presented in this paper are not publicly available at this time but may be obtained from the authors upon reasonable request.

\bibliography{Reference_apc_2022}






\end{document}